\documentclass[12pt]{article}

\usepackage{calrsfs}
\DeclareMathAlphabet{\pazocal}{OMS}{zplm}{m}{n}
\usepackage[letterpaper, nohead,dvips]{geometry}
\usepackage{amsmath}
\usepackage{amsfonts}
\usepackage{amssymb}
\usepackage{enumerate}
\usepackage{amsthm}
\usepackage{setspace}
\usepackage[all,cmtip]{xy}
\usepackage{graphicx}
\usepackage{chngcntr}
\usepackage[english]{babel}
\usepackage[section]{placeins}
\usepackage{float}
\usepackage{tikz-cd}
\usepackage{subfig}
\usepackage[latin10,utf8]{inputenc}

\floatstyle{boxed}
\restylefloat{figure}

\theoremstyle{plain}

\theoremstyle{definition}

\numberwithin{equation}{section}

\begin{document}

\title{The Hamiltonian approach to the problem of derivation of production functions in economic growth theory}

\author{Roman G. Smirnov\footnote{e-mail:  Roman.Smirnov@dal.ca}\,   and Kunpeng Wang\footnote{e-mail: kunpengwang@dal.ca} \\Department of
  Mathematics and Statistics\\
Dalhousie University\\ Halifax, Nova Scotia, Canada
  B3H~3J5}

\maketitle

\begin{abstract}
We introduce  a general Hamiltonian framework that appears to be a natural setting for the derivation of  various production functions in economic growth theory, starting with the celebrated Cobb-Douglas function. Employing our method, we investigate some existing  models and propose a new one as special cases of the  general $n$-dimensional Lotka-Volterra system of eco-dynamics. 
\end{abstract}

\section{Introduction}
\label{s1}
As is well known, both mathematicians  and economists   have been using various  methods to derive   production functions in economic growth theory ranging from data analysis \cite{CD28, PHD76} to symmetry and Lie group theory methods \cite{Sato81, SR14}. In this paper we will enlarge the set of  available tools by incorporating a Hamiltonian formalism  into the theory.

To show the workings of the Hamiltonian formalism in the study of technical progress and production functions in economics, let us recall first that symmetry methods have already proven to be a very powerful tool in this context, which was  demonstrated by Sato \cite{Sato81} (see also  Sato and Ramachandran \cite{SR14}, the relevant references in \cite{SW19}, and, for example, Perets and Yashiv \cite{PY18}). In particular, the authors in \cite{SW19} have extended Sato's approach to derive a new family of production functions under the assumption of logistic growth in factors. 

It is our contention that the theory can be further developed at this point  by recasting its setting within a Hamiltonian framework. More specifically, we will redefine the existing models \cite{Sato81,  SW19} and introduce a new one by presenting them as special cases of the general $n$-dimensional Lotka-Volterra model in population dynamics (see, for example, Kerner \cite{EK72} and the relevant references therein). This model is given by the following formula:  
\begin{equation}\label{LV1}
\dot{x}_i=x_i\left(b_i+\sum_{j=1}^n a_{ij}x_j\right), \quad i=1,\ldots,n,
\end{equation}
where the linear terms describe the Malthusian growth (or decay) of the species in question $x_1, \ldots, x_n$ in the absence of interaction (i.e., when the parameters $a_{ij}$ all vanish), while the quadratic terms tell us about the binary interraction between the species, assuming spatial homogeneity.  More specifically, $a_{ij}= \frac{1}{\beta_i}\alpha_{ij}$, where $\beta_i$ is Volterra's ``equivalent number" parameter that has the meaning of mean effective biomass of the individuals in the $i$th species, while $\alpha_{ij}$ is normally assumed to be  a  skew-symmetric matrix representing the interaction strength of species $i$ with species $j$ \cite{EK72}. We recall that the Lotka-Volterra systems with vanishing linear terms (i.e., when $b_i = 0$, $i=1, \ldots, n$ in (\ref{LV1})), as well as their integrable perturbations are an important  and well-studied topic in the  field of mathematical physics, in particular, they appear as discretizations of the KdV equation  (see, for example, Bogoyavlenskij {\em et al} \cite{BIY07}  and Damianou {\em et al} \cite{DEKV17} for more details and references). Furthermore, Plank \cite{MP95, MP96} (see also Kerner \cite{EK96}) studied general $n$-dimensional Lotra-Volterra systems from the Hamiltonian viewpoint and  found bi-Hamiltonian formulations for particular 3-dimensional model (\ref{LV1}).

In what follows we will adapt some of these results to the problem of the determination of a production function under the assumption of a holothetic growth in factors. 

\section{The Hamiltonian formalism via Poisson geometry} 
\label{s2} 

All models that we discuss in this paper will be studied  within the framework of the Hamiltonian systems defined on Poisson manifolds. Recall  that a {\em Poisson structure} on a manifold $M$ is a skew-symmetric ${\mathbb R}$-bilinear bracket 
$$\{\cdot, \cdot\}: C^{\infty}\times C^{\infty} \rightarrow C^{\infty},$$
satisfying the Leibnitz rule
$$\{f, gh\} = \{f,g\}h + g\{f, h\}$$ 
and the Jacobi identity
$$\{f,\{g,h\}\} + \{h,\{f,g\}\} + \{g,\{h,f\}\} =0,$$
for all $f,g, h \in C^{\infty}(M)$. The pair of a manifold $M$ and a Poisson structure defined on $M$ is called a {\em Poisson manifold}. 
Next, let $(M, \{\cdot, \cdot \})$ be a Poisson manifold, then  the  vector field $X_f$ given by 
$$X_f = \{f, \cdot\}$$
is called the {\em  Hamiltonian vector field} determined by the {\em Hamiltonian function} $f$. Note that  the value of $\{f, g\}$  at any point $p \in M$ depends linearly  on the differentials $\mbox{d}f, \mbox{d}g$ at $p \in M$. In this view, the bracket $\{\cdot, \cdot \}$ gives rise to a Poisson bi-vector field $\pi \in \mathfrak{X}^2 (M)$ $= \Gamma (\Lambda^2TM)$ such that 
$$\pi (\mbox{d}f, \mbox{d}g) = \{f, g\},$$
for all $f, g \in C^{\infty}(M)$. Conversely, given a  Poisson bi-vector $\pi \in \Gamma (\Lambda^2TM)$, then $\pi$ defines the corresponding Poisson bracket satisfying the properties specified above. In what follows we will refer to a Poisson manifold as a pair $(M, \pi)$, which gives rise to the following definition of a Hamiltonian vector field
\begin{equation}
\label{Ham1}
X_f = \pi df,
\end{equation}
or, in terms of local coordinates $(x_1, \ldots, x_n)$ near $p \in M$, 
\begin{equation}
\label{Ham2}
X^i_f = \pi^{i\ell}\frac{\partial f}{\partial x_{\ell}},
\end{equation}
where $X_f = (X^1_f, \ldots, X^n_f)$ and $n = \dim M$. See, for example, Fernandes and M\u{a}rcu\textcommabelow{t} \cite{FM15} for more details. We note, however, that most of the above formulas can be represented in a uniform way via the Schouten bracket $[\cdot, \cdot ]$ \cite{S40}. Thus, for instance, the Jacobi identity condition for a Poisson bracket $\{\cdot, \cdot\}$ defined by a Poisson bi-vector $\pi$ is simply equivalent to the condition $[\pi, \pi] = 0$. A Hamiltonian vector field $X_f$ defined on a Poisson manifold $(M, \pi)$ can be now determined as 
$$X_f = [\pi, f].$$
Similarly, the Poisson bracket of  any functions $f, g \in C^{\infty} (M)$ defined on a Poisson manifold $(M, \pi)$  may be defined via the Schouten bracket as 
$$\{f, g\} = [[\pi, f],g]$$
and so on (see, for example,  \cite{RS97} for more details).  We note that the Hamiltonian formalism determined  within  the framework of Poisson geometry  is a preferred choice for the investigation of the models related to the Lotka-Volterra system. Indeed,  when the number of species $n$ in (\ref{LV1})  is odd,  it is not possible to employ the Hamiltonian formalism via  symplectic geometry, since a symplectic form is by definition non-degenerate, while a Poisson bi-vector $\pi$ defined above may be degenerate. 

\section{Sato's model}
\label{s2}

Recall, Sato \cite{Sato81} (see also pp. 4-5 in \cite{SW19}) employed Lie group theory methods to derive the Cobb-Douglas function as an invariant of a one parameter group action determined by exponential growth in  labor, capital,  and production. His other goal was to resolve the Solow-Stigler controversy \cite{Sato81, Solow56} that can be described as the  observation that the increase of output is not proportional to the growth of labor and  capital  in the production function, which,  in turn,  implies   that the technical progress  should also be taken into the account in the development of any growth model. The assumption that labor, capital,  and production grow exponentially  leads to a simple dynamical system, which we can interpret as a special case of the system (\ref{LV1}).  Indeed,   in (\ref{LV1}) let $n=3$   and the parameters $a_{ij}$ all vanish. Then,  we have 
\begin{equation}
\label{model1}
\dot{x}_i=b_ix_i, \quad i = 1,2,3,
\end{equation}
where $x_1 = L$ (labor), $x_2 = K$ (capital), $x_3 = f$ (production), $b_1 = b$, $b_2 = a$ and $b_3 = 1$ in Sato's notations. Next, employing the approach presented in \cite{MP96} {\em mutatis mutandis} (see also \cite{BR91, DP14} for more details), we rewrite (\ref{model1}) as the following Hamiltonian system: 
\begin{equation}
\label{model1HS}
\dot{x}_i = X^i_{H}= \pi_1^{i\ell }\frac{\partial H
}{\partial x_{\ell}}, 	\quad i = 1,2,3.
\end{equation}
Here 
\begin{equation}
\label{Poisson1}
\pi = - x_ix_j\frac{\partial}{\partial x_i}\wedge \frac{\partial} {\partial x_j}, \quad i, j = 1,2,3
\end{equation}
is the quadratic  (degenerate) Poisson bi-vector that defines the Hamiltonian function
\begin{equation}
\label{model1Hamiltonian}
H = \sum_{k=1}^3c_k\ln x_k
\end{equation}
via $X_{H} = [\pi, H]$, in which  the parameters $c_k$ are solutions to the rank 2 algebraic system $A {\bf c} = {\bf b}$ determined by  the skew-symmetric  $3 \times 3$ matrix $A$
$$
A = \begin{bmatrix} 0 & -1 & -1 \\ 
1 & 0 & - 1\\ 
1 & 1 & 0
\end{bmatrix}, 
$$ 
${\bf c} = [c_1, c_2, c_3]^{T}$ with all $c_k>0$, and ${\bf b} = [b_1, b_2, b_3]^T$, satisfying the condition 
\begin{equation}
b_1 + b_3 = b_2.
\label{c1}
\end{equation}
We observe next that
$\mbox{div} X_H = b_1 + b_2 + b_3$, which implies that the Hamiltonian vector field $X_H$ defined above is not incompressible. In particular, if all of the ``growth" parameters $b_1$, $b_2$, $b_3 >0$, the economy, interpreted as a volume-element, grows. We also note that the vector field $X_H$ is irrotational, i.e., $\nabla \times X_H = 0$, that is $X_H$ is a conservative vector field. Indeed, $X_H = \nabla f$, where $f (x_1, x_2, x_3) = \frac{1}{2}(b_1 x_1^2 + b_2 x_2^2 + b_3 x_3^2).$

Alternatively, we can introduce the following new variables 
\begin{equation}
\label{Sub1}
v_i = \ln x_i, \quad i = 1, 2,3,
\end{equation}
 which lead to an even simpler form of the system (\ref{model1}), namely
\begin{equation}
\label{model1-2}
\dot{v}_i = b_i, \quad i = 1, 2, 3. 
\end{equation}
Interestingly, the substitution (\ref{Sub1}) is exactly the one used by Cobb and Douglas in \cite{CD28} to derive the celebrated Cobb-Douglas production function (\ref{CDF}). More specifically, they  graphed the three functions $v_1$, $v_2$ and $v_3$, coming from specific data (i.e., specific variables representing ``labor force" $L$, ``fixed capital" $K$,  and ``physical product" $f$ determined by E. E. Day's index of the physical volume of the US production for the years 1899-1922), and observed the correlations that led to the introduction of the function (\ref{CDF}) for $A = 1.01$, $\alpha = 3/4$, and $\beta  = 1/4$. 

Note that (\ref{model1-2}) is also a Hamiltonian system, provided $b_1 + b_3 = b_2$, defined by  the corresponding (degenerate) Poission bi-vector $\tilde{\pi}$ with  components $$\tilde{\pi}^{ij} = - \frac{\partial}{\partial v_i} \wedge \frac{\partial}{\partial v_j}$$ and  the corresponding  Hamiltonian $$\tilde{H} = \sum_{k=1}^3c_k v_k.$$ 

Observing that the  function  $H$ given  by (\ref{model1Hamiltonian}) is a constant of the motion of the Hamiltonian system (\ref{model1HS}), and then  solving the equation $ \sum_{k=1}^3c_k\ln x_k  = H =  \mbox{const}$ for $x_3$, we arrive at the  celebrated Cobb-Douglas production function
\begin{equation}
\label{CDF}
Y = f(L, K) = A L^{\alpha} K^{\beta},
\end{equation} 
 after the identification $x_1 = L$, $x_2 = K$, $x_3 = f$, $A = \exp\left(\frac{H_1}{c_3}\right)$, $\alpha = -\frac{c_1}{c_3}$, $\beta  = -  \frac{c_2}{c_3}$. Note that the elastisities of substitution $\alpha$ and $\beta$, satisfying the condition 
\begin{equation}
\label{CRS}
\alpha + \beta = 1,
\end{equation}
assure that the production function enjoys constant return to scale. In order to make sure that the Cobb-Douglas function (\ref{CDF}) derived analytically under the assumption that  labor, capital, and production grow exponentially satisfies the condition (\ref{CRS}), Sato \cite{Sato81} introduced  the ``simultaneous holotheticity condition", from which an aggregate production function (\ref{CDF}) satisfying (\ref{CRS}) could be derived. Mathematically, this condition is equivalent to the existence of  a two-dimensional integrable distribution that represents simultaneous technical change in two sectors of economy that are characterized by the same aggregate production function. We will use the bi-Hamiltonian approach (see, for example, \cite{MP96, RS97} for more details; to learn more about the origins of the theory  --- see  \cite{PS05}) to derive the Cobb-Douglas function (\ref{CDF}), satisfying (\ref{CRS}). Following the bi-Hamiltonian treatment of  three-dimensional Lotka-Volterra systems presented in  \cite{MP96}, we introduce the following bi-Hamiltonian structure for the dynamical system (\ref{model1}): 
\begin{equation}
\label{BHS1}
\dot{x}_i = X_{H_1, H_2} = [\pi_1, H_1] = [\pi_2, H_2], \quad i = 1, 2, 3, 
\end{equation} 
where the Hamiltonian functions $H_1$ and $H_1$ are given by
\begin{equation}
\label{H1}
H_1 = b\ln x_1 + \ln x_2 + a \ln x_3, 
\end{equation}

\begin{equation}
\label{H2}
H_2 = \ln x_1 + a\ln x_2 + b \ln x_3.
\end{equation} 

The Hamiltonian functions $H_1$ and $H_2$ correspond to the Poisson bi-vectors $\pi_1$ and $\pi_2$ 
\begin{equation}
\pi_1 =  a_{ij} x_ix_j\frac{\partial}{\partial x_i}\wedge \frac{\partial} {\partial x_j}, \quad i, j = 1,2,3
\label{p1}
\end{equation}
and
\begin{equation}
\pi_2 =  b_{ij} x_ix_j\frac{\partial}{\partial x_i}\wedge \frac{\partial} {\partial x_j}, \quad i, j = 1,2,3
\label{p2}
\end{equation}
respectively under the conditions
\begin{equation} 
\label{conditions}
\left\{\begin{array}{rcl}
bb_1 + b_2 + ab_3 & = &0, \\ b_1 + ab_2+b_3 b & = & 0. 
\end{array}\right.
\end{equation} 
Note the conditions (\ref{conditions}) (compare them to (\ref{c1})) assure that $\pi_1$ and $\pi_2$ are indeed Poisson bi-vectors compatible with the dynamics of (\ref{model1}) and corresponding to the Hamiltonians (\ref{H1}) and (\ref{H2}) respectively. 
Solving the linear system (\ref{conditions}) for $a$ and $b$ under the additional condition $b_1b_2 - b_3^2 \not=0$, we arrive at 
\begin{equation}
\label{ab}
a = \frac{b_2b_3 - b_1^2}{b_1b_2 - b_3^2}, \quad b = \frac{b_1b_3 - b_2^2}{b_1b_2 - b_3^2}.
\end{equation}
Consider now the first integral $H_3$ given by 
\begin{equation}
H_3 = H_1 - H_2 = (b-1)\ln x_1 + (1-a)\ln x_2 + (a-b)\ln x_3. 
\label{H3}
\end{equation}
Solving the equation (\ref{H3}) for $x_3$, we arrive at the Cobb-Douglas function (\ref{CDF}) with the elastisities of substitution  $\alpha$ and $\beta$ given by
$$\alpha  = \frac{a-1}{a-b}, \quad \beta = \frac{1-b}{a-b},$$
where $a$ and $b$ are given by (\ref{ab}). Note $\alpha + \beta = 1$,  as expected. Also, $\alpha, \beta >0$ under the additonal condition $b_2>b_3>b_1$, which implies by (\ref{model1}) that capital ($x_2 = K$) grows faster than production ($x_3 = f$), which, in turn,  grows faster than labor ($x_1 = L$).  We conclude, therefore, that the existence of a bi-Hamiltonian structure was crucial for our considerations, because it  enabled us to produce a unique pair of acceptable, from the economic viewpoint elasticities of substitution  $\alpha$ and $\beta$  in (\ref{CDF}) directly from the parameters $b_1, b_2$ and $b_3$ that determined the dynamics of the system  (\ref{model1}).

\section{The logistic growth model}
\label{s3}

Recall that the authors extended Sato's approach  in \cite{SW19},  replacing the assumption about exponential growth in labor, capital  and production with the corresponding assumption that  labor, capital, and production grow logistically, arriving, as a result, at the following dynamical system
\begin{equation}
\label{model2}
\dot{x}_i = b_ix_i\left(1 - \frac{x_i}{N_i}\right), \quad i = 1, 2, 3, 
\end{equation}  
where $x_1 = L$ (labor), $x_2 = K$ (capital), $x_3 = f$ (production), $b_1 = b$, $b_2 = a$, and $b_3 = 1$ in Sato's notations, adopted in \cite{SW19},  and the parameters $N_i$ denote the corresponding carrying capacities. Furthermore, we  employed Sato's approach \cite{Sato81} to integrate this dynamical system and thus  derive a new production function (see (4.5) in \cite{SW19}).  We will now use   the  Hamiltonian approach by treating the system (\ref{model2}) as a particular case of the general Lotka-Volterra system (\ref{LV1}) and, therefore, a Hamiltonian system as such. Indeed, we first note that in this case $n = 3$ and the $3\times 3$ matrix determined by the parameters $a_{ij}$ in (\ref{LV1})  is diagonal, with $a_{ii} = - \frac{b_i}{N_i}$. Following \cite{EK96}, we rewrite the system (\ref{model2}) in terms of the new variables given by $x_i = N_i e^{v_i}$, $i = 1,2,3$, which yields  
\begin{equation}
\label{model2-2}
\dot{v}_i = b_i (1-e^{v_i}), \quad i = 1, 2, 3. 
\end{equation}
We also introduce the (degenerate) Poisson bi-vector $\pi_3$ 
\begin{equation}
\pi_3 = -(1-e^{v_i})(1- e^{v_j})\frac{\partial }{\partial  v_i}\wedge \frac{\partial}{\partial v_j}, \quad i, j  = 1,2,3.
\end{equation}
Note, the $(2,0)$-tensor $\pi_3$  is skew-symmetric and satisfies the condition $[\pi_3, \pi_3] = 0$, where $[\cdot, \cdot]$ denotes the Schouten bracket \cite{S40}. The corresponding Hamiltonian function $H_3$, satisfying the equation 
$$\dot{v}_i =V^i_{\tilde{H}_3} = \pi_3^{i\ell }\frac{\partial \tilde{H}_3}{\partial v_{\ell}}, 	\quad i = 1,2,3,$$ is found to be
\begin{equation}
\tilde{H}_3 = \sum_{k=1}^3c_k\left(v_k - \ln(1- e^{v_k})\right), 
\end{equation} 
or,  in terms of the original variables,  $$H_3 = \sum_{k=1}^3c_k\ln\frac{x_k}{|N_k - x_k|}.$$ Next, solving the equation $$\sum_{k=1}^3c_k\ln\frac{x_k}{|N_k - x_k|} = H_3 = \mbox{const}$$ for $x_3$ and indentifying $x_1 = L$, $x_2 = K$, $x_3 = f$, $N_1 = N_L$, $N_2 = N_K$, $N_3 = N_f$,  $-\frac{c_1}{c_3} = \alpha$, $-\frac{c_2}{c_3} = \beta$, $e^{-H_2/c_3} = C$, we arrive at the production function
\begin{equation}
\label{newfunction}
Y = f(L, K) = \frac{N_fL^{\alpha}K^{\beta}}{C|N_L-L|^{\alpha}|N_K- K|^{\beta} + L^{\alpha}K^{\beta}} 
\end{equation}
derived  in \cite{SW19}. Recall that a similar, ``S-shaped"  production function 
\begin{equation}
\label{Sshaped}
 Y = g (L, K) = \frac{aL^pK^{1-p}}{1+bL^pK^{1-p}}
\end{equation}
was recently introduced, employing a heuristic approach, see   \cite{ACKL13, CEL12, EBC14, LLM15} and the relevant references therein for more details and applications. Note that the production function (\ref{Sshaped}) is reducible to the Cobb-Douglas function (\ref{CDF}) (i.e., when $b = 0$). Also, we observe that the new production function  (\ref{newfunction}) is reducible to the production function (\ref{Sshaped})  when $K$ and $L$ $\ll$ $N_K$ and $N_L$ respectively, $N_L, N_K \approx 1$, $C = 1$ in (\ref{newfunction}) and $a = N_{f}$,   $b = 1$ in (\ref{Sshaped}) .

\section{A new model involving debt}
\label{s4} 

In what follows, we  employ  the well-established variables used in modelling   processes in  economic growth theory, namely $L$ (labor),  $K$ (capital),  $f$ (production),  and we introduce in addition, a new variable $D$ (debt). Note that treating debt as an independent variable is a novel but already acceptable practice in economic modelling (see, for example, \cite{AJ13}).  We shall assume that debt and capital interact in a way similar to the predator-pray collisions in eco-dynamics \cite{EK72}. More specifically,  we assume that in the absence of capital, debt grows exponentially and vice versa --- when debt is absent,  capital also grows exponentially. At the same time, more capital can  ``eat" debt (i.e., the debt gets paid off), while more debt diminishes by the same token any disposable income (capital). As for production and labor, we shall assume they grow logistically. Therefore, we consider  the following $4$-dimensional dynamical system: 
\begin{equation} 
\label{model3}
\begin{array}{rcl}
\dot{x}_1& = & x_1(b_1 + a_{12}x_2), \\[0.2cm]
\dot{x}_2& = & x_2(b_2 + a_{21}x_1),\\[0.2cm]
\dot{x}_3& = & x_3b_3\left(1 - \frac{x_3}{N_3}\right), \\[0.2cm]
\dot{x}_4& = & x_4b_4\left(1 - \frac{x_4}{N_4}\right), \\ [0.2cm] 
\end{array}
\end{equation}
where $x_1 = K$, $x_2 = D$, $x_3 = f$, $x_4 = L$, $N_3 = N_f$, $N_4 = N_L$. Not that the  system (\ref{model3}) is also a special case of the $n$-dimensional Lotka-Volterra model  (\ref{LV1}). Changing the variables $v_1 = \ln \left(-\frac{a_{12}}{b_1} x_1\right)$, $v_2 =  \ln \left(-\frac{a_{21}}{b_2} x_2\right)$, $v_3 = \ln \frac{x_3}{N_3},$ $v_4 = \ln \frac{x_4}{N_4}$, and assuming $a_{12}b_1, a_{21}b_2 <0$,  yields the Hamiltonian system 
\begin{equation}
\label{H4}
v_i = V^i_{H_4} =  \pi_4^{i\ell} \frac{\partial H_4}{\partial v_{\ell}}, \quad i = 1, 2, 3, 4,
\end{equation}
 where the Hamiltonian function is given by 
\begin{equation}
\begin{array}{rcl}
H_4 & =&  b_1(v_2 - e^{v_2}) - b_2(v_1- e^{v_1}) + \\[0.5cm]
& & \displaystyle \frac{1}{b_3}\left(v_3 - \ln (1- e^{v_3})\right) - \frac{1}{b_4}\left(v_4 - \ln (1-e^{v_4})\right),

\end{array}
\label{model3-H3}
\end{equation}
corresponding to the Poisson bi-vector 
\begin{equation}
\label{P4}
\pi_4 = \pi^{ij}_4\frac{\partial}{\partial v_i} \wedge \frac{\partial }{\partial v_j},
\end{equation}
 with non-zero components:  $\pi_4^{12} = - \pi_4^{21} = -1$, $\pi_4^{34} = -\pi_4^{43} = - b_3b_4(1-e^{v_3})(1-e^{v_4})$ (All other components vanish).  Solving the  equation (\ref{model3-H3}) for $f$, we arrive at the following new production function that accounts for interactions between $K$ (capital) and $D$ (debt)
\begin{equation} 
\label{newfunction1}
Y= f(L, K, D) = \frac{N_f e^{b_3G(L, K, D)}}{1 + e^{b_3G(L, K, D)}},
\end{equation} 
where the fuction $G$ is given by 
$$\begin{array}{rcl} \displaystyle G & = &   \displaystyle C - b_1\left[\ln\left(-\frac{a_{21}}{b_2}D\right) + \frac{a_{21}}{b_2}D\right]  + b_2\left[\ln\left(-\frac{a_{12}}{b_1}K\right) + \frac{a_{12}}{b_1}K\right] +\\[0.5cm] 
& &\displaystyle  \frac{1}{b_4}\ln \frac{L}{N_L - L}, \quad C \in \mathbb{R}.
\end{array}
$$

\section{Concluding remarks} 
\label{s5} 

Mathematicians often say that ``a mathematical problem is  essentially solved when it is reduced to an algebraic problem." In this paper the authors  have  reduced several  problems of the derivation of a production function  to the corresponding  algebraic problems by employing the Hamiltonian approach and describing the dynamics in question in each case as a special case of the Lotka-Volterra model (\ref{LV1}). In particular,  we have rederived the celebrated Cobb-Douglas production function (\ref{CDF}) with economically acceptable elasticities of substitution as a linear combination of two Hamiltonians of the bi-Hamiltonian structure (\ref{BHS1}) defined by two quadratic (degenerate) Poisson bi-vectors. Our next model came from a recent paper \cite{SW19} in which we extended Sato's ideas coming from the Lie group theoretical framework for the theory of endogeneous technical progress developed in \cite{Sato81} and \cite{SR14} by assuming logistic rather than exponential growth in factors. In this case too, we identified the corresponding dynamical system as a special case of the Lotka-Volterra model (\ref{LV1}) and a Hamiltonian system as such, which enabled us to derive the corresponding production function (\ref{newfunction}) as a Hamiltonian. The last model presented in this paper is new --- we have introduced an additional variable (debt) and described the dynamics built around the ``predator-prey" type interaction between capital and debt also as a special case of the Lotka-Volterra model (\ref{LV1}), which ultimately led to the derivation of a new production function (\ref{newfunction1}).  

Further analysis of the models studied in this paper, as well as their generalizations, using quantitative and qualitative research methods, including the methods of statistical mechanics, will be presented in a forthcoming article by the authors.

\end{document}